# Raman studies of suspensions and solutions of singlewall carbon nanotubes

N. Izard[1], A. Pénicaud[2], E. Anglaret[1]
1-Groupe de Dynamique des Phases Condensées, UMR CNRS 5581,
Université Montpellier 2, Montpellier, France
2-Centre de Recherche Paul Pascal, CNRS-Université Bordeaux 1,
Pessac, France

**ABSTRACT**
Raman spectroscopy is used to probe the structure and electronic properties of nanotubes dispersed in a liquid phase. We show that the radial breathing modes are upshifted in suspensions due to the molecular pressure of the solvent. On the other hand, we directly probe charge transfer in solutions of nanotube polyelectrolytes and its reversibility after oxydation in air.

**INTRODUCTION**
Homogeneous dispersions of single wall nanotubes in liquids are required for optical applications and processing in materials science. Functionnalization leads to changes in the structure of the nanotubes [1]. An alternative and popular route is the preparation of aqueous suspensions with the help of surfactants [2]. On the other hand, a new route was recently explored to prepare solutions of nanotubes, via chemical protonation [3] or chemical reduction in polar organic solvents [4].
Raman spectroscopy is the most popular technique to characterize samples of single wall carbon nanotubes (SWNT), because it provides reliable informations not only on their vibrational properties, but also on their structural and electronic properties [5,6]. Raman is also an effective technique to study nanotubes dispersed in liquids. However, one must be aware of possible new structural interactions and changes in the electronic properties due to a new physical and chemical environment.
In this paper, we study the Raman signatures of suspensions and solutions of SWNT. We especially focus on the changes in the radial breathing mode frequency between powders and suspensions. We also investigate charge transfers in the liquid phase for nanotube polyelectrolyte solutions.

**EXPERIMENTAL**

Single wall carbon nanotube samples were prepared by the electric arc technique (from GDPC and Nanoledge, Inc.). Aqueous suspensions were prepared using ionic (SDS) or anionic (Triton X100) surfactants. After a few hours of low-power sonication, homogeneous and stable suspensions of bundles are obtained, as stated by electron microscopy and Raman measurements (not shown). By contrast, high power sonication (500 W during 15 minutes) leads to exfoliation of the bundles [7]. After an ultracentrifugation step (120000 g during 4 hours), aqueous suspensions of individual SWNT are formed, as stated by electron microscopy, photoluminescence [8] and photon correlation spectroscopy [9] techniques.
Solutions of reduced SWNT (anionic form) were prepared from Na salts [10]. Indeed, it was shown recently that Na(THF)SWNT salts dissolve spontaneously in DMSO and other polar solvents to form a nanotube polyelectrolyte solution [4]. Since such solutions are not stable in air, we studied them both in sealed cells and after oxydation in air.
The Raman experiments were carried out on a Jobin-Yvon spectrometer, using the green (2.41 eV) and red (1.92 eV) lines of an Ar-Kr laser, and on a Bruker RFS100 FT spectrometer using the 1.16 eV excitation line of a Nd-YAG laser.

## RESULTS AND DISCUSSION

### Frequency of the radial breathing modes

We first present a compared study of the radial breathing modes (RBM) in powders and suspensions. Measurements of the RBM frequency is widely used to estimate the diameter of SWNT. Indeed, the RBM frequency is expected to be inversely proportional to diameter for *isolated* nanotubes [5,6]. However, in most of the samples, nanotubes are not isolated. They usually assemble into crystallines nano-bundles, where the RBM upshift due to intertube interactions [5,6]. Even when experiments are carried out on *individual* nanotubes, interactions with the substrate must be considered to determine the relation between RBM frequency and tube diameter [11]. Several Raman studies were performed on suspensions or solutions [3,12,13] but the effect of the interaction with the solvent on the RBM frequencies was not addressed so far. However, it was shown that the internal pressure of the solvent induce a shift of the second harmonic of the D-band, the so-called D* (or G') band, which also involves radial motions of the carbon atoms [13].

In figure 1, we compare the RBM bunches for powders, suspensions of bundles (SB) and suspensions of individual SWNT (SIS) made from the *same* electric arc samples. The most striking feature is a significant and *global* upshift of the RBM bunch from powder to suspensions, independently of the aggregation state of the nanotubes. In order to describe more precisely this upshift, each spectrum was fitted with a sum of lorentzians (fig. 1). Details can be found elsewhere [8]. Good fits were obtained when allowing the frequencies to vary. In this case, systematic, global upshifts of the main Raman lines were found. By contrast, the fits were poor when fixing the line frequencies and fitting the intensities only. The results of the fits are summarized in figure 2. In this diameter range, the shift is almost constant, of about 5 to 6 cm$^{-1}$ (note that for smaller tube diameters, the upshift was found to be significantly smaller [8]), independently of RBM frequency or laser energy (and therefore, independently of the semiconducting or metallic properties of the tubes).

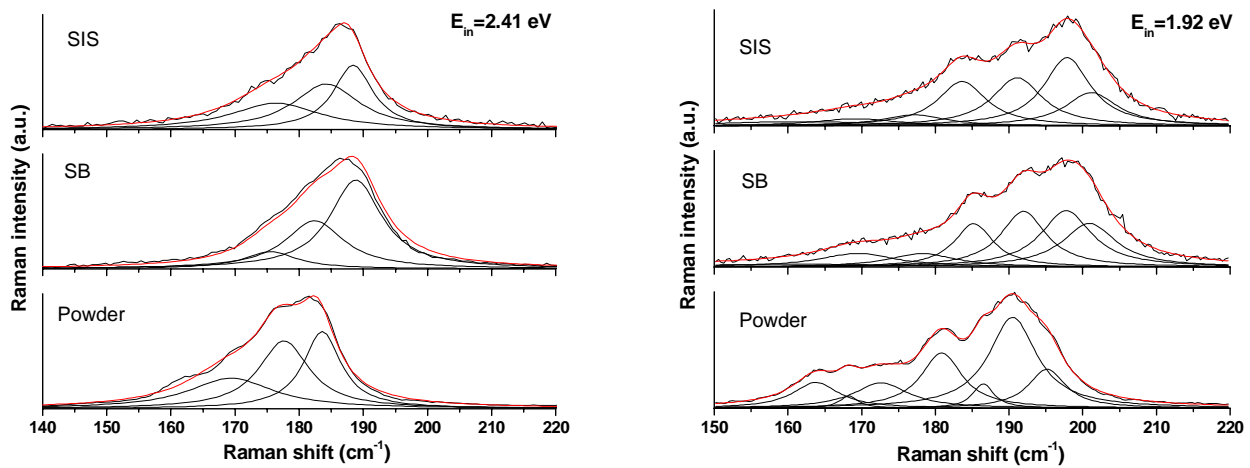

**Figure 1** : Compared RBM spectra for powders, suspension of bundles (SB) and suspension of individual SWNT (SIS), for laser energies 2.41 eV (left) and 1.92 eV (right). A linear background was substracted from the spectra for clarity.

There are two possible explanations for such changes in the RBM frequencies. First, one may argue that the electronic properties of the nanotubes change when their environment change. Therefore, the resonant Raman effect may select different nanotubes for the same samples in powders and in suspensions. Since the energy of the allowed optical transitions (AOT) of SWNT are, in first approximation, inversely proportional to the tube diameter [14,15], a selection of smaller diameters would mean a decrease (redshift) of the AOT. Calculations are missing to support this hypothesis, and one rather expects an increase (blueshift) of the AOT due to hydrostatic pressure [15]. Anyway, even if it can not be ruled out that changes in the electronic properties may modulate the RBM intensities, we claim that they can not be responsible for a *systematic* upshift. Indeed, the energy of the AOT depends not only on tube diameter but also on chiral angle and therefore the dispersion of the AOT is rather large for a given diameter [14,15]. Thus, changes in the optical properties of the tubes would likely modulate the relative intensities of the RBM and perhaps make some peaks appear or disappear, but not in the systematic way observed in the experiments.

Therefore, we conclude that the global upshift of the RBM frequency must be attributed to structural interactions between nanotubes and the surrounding solvent molecules, *i.e.* to the molecular pressure of the solvent. Wood and Wagner also reported an upshift of the D* band when nanotubes are dispersed in liquids [13]. They showed that the upshift is proportional to the cohesive energy density (CED) of the liquid, *i.e.* to its internal pressure [13]. Therefore it is tempting to compare the shift observed in suspensions to that measured under hydrostatic pressure. A shift of 5 to 6 $cm^{-1}$ corresponds to an hydrostatic pressure of a few hundred MPa [16], which is of the same order of magnitude of the CED of organic liquids. This similitude give additional credit to our analysis. However, the comparison must be made with care since in the case of hydrostatic pressure, a shift of the tangential modes (TM) is also observed [16]. By contrast, in the suspensions, a weak but systematic downshift of 1 to 2 $cm^{-1}$ was observed [8]. The interpretation of this downshift still remains an open question.

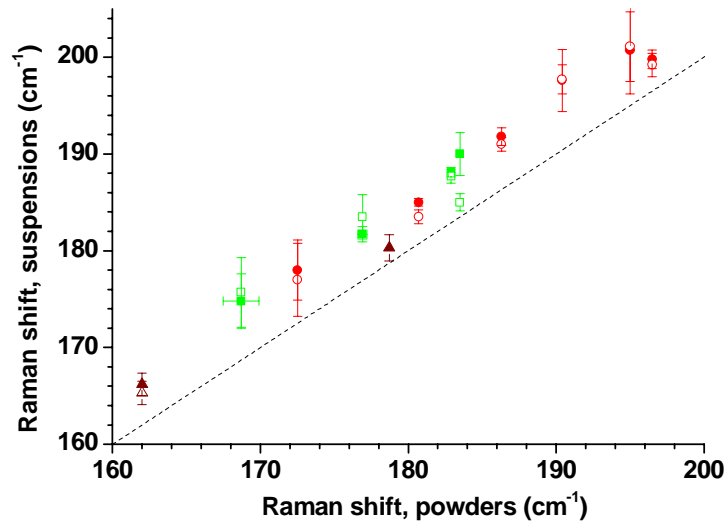

**Figure 2** : RBM Raman shift in the suspensions with respect to shift in the powders. The solid and open symbols are for SB and SIS, respectively. The squares, circles and triangles are for laser energies 2.41, 1.92, and 1.16 eV, respectively. The dashed line is a guide for the eye.

On the other hand, no significant change in the Raman spectra were observed after exfoliation in the suspensions. The only striking change is a strong photoluminescence signal in the near infrared which confirms exfoliation in SIS (not shown) [7,8]. This result can seem surprising at first sight since calculations predict a downshift of about 10 cm$^{-1}$ for isolated tubes with respect to bundles [5,6]. However, we point out that SWNT in suspensions can not be considered as *isolated* tubes, because they are in contact with the solvent molecules around. We believe that nanotube-solvent interactions explain why no RBM shift is observed between SB and SIS.

**Evidence of charge transfers in SWNT polyelectrolyte solutions**

We now focus on the Raman signatures of solutions of SWNT salts. Doping of nanotubes with alkali atoms has been widely investigated in the solid state [10,18,19]. SWNT can be doped in the vapor phase [18] or via chemical routes [19]. The Raman signature of alkali-doped SWNT was found to be universal, independently of the nature of the alkali-atoms and of the doping technique [19]. It was recently shown that Na or Li salts of nanotubes are soluble in DMSO and other polar solvents, for concentrations up to 2 mg.g$^{-1}$ [4].

Typical Raman spectra for the solutions are presented in figure 3, and compared to the spectra of powder and suspensions. The signature of the doped phases appears both in the ranges of the radial breathing modes (RBM) and of the tangential modes (TM). For pristine (undoped) samples, the spectral profile of the TM changes with the excitation line. The spectra for powders measured at 2.41 eV display symmetric and narrow lorentzian-like lines (fig. 3, top). This signature corresponds to semi-conducting nanotubes of about 1.3-1.4 nm, resonantly excited at 2.41 eV. By contrast, the spectra measured at 1.92 eV display an asymetric and broad profile (fig. 3, bottom). This was assigned to Breit-Wigner-Fano resonance for metallic tubes of about 1.3-1.4 nm, resonantly excited at 1.92 eV [5,6]. The TM spectra of suspensions are very close to those of powders [19]. At 2.41 eV, the profile is the same (except a slight downshift). At 1.92 eV, the profile is slightly different than that of the powder but it remains broad, asymmetric, BWF-like. By contrast, the TM spectra of solutions is very different. It is featured by a strong decrease of intensity, by an upshift of the main component and by a loss of the BWF profile at 1.92 eV. All these features are also observed for solid Na-NT salts [18]. The loss of intensity and the similarity of the spectra at 2.41 and 1.92 eV are characteristic of non-resonant Raman spectra. Indeed, optical transitions in the visible vanish upon doping [10]. On the other hand, the upshift of the TM was observed in the first steps of vapor phase doping, and in solid doped samples prepared by the chemical route, and assigned to the formation of a stable doped phase [18]. The loss of Raman intensity due to the loss of resonant conditions is even more drastic in the RBM range : no Raman signal can be detected in the solutions. Finally, the solubilisation of the anionic form of SWNT is definitively confirmed by the spectra measured after oxydation in air (top spectra in figure 3). Both the RBM and TM profiles are very close to those of suspensions. They are featured by a frank upshift of the RBM with respect to powders, and a slight downshift of the TM, as discussed in the first part of this paper.

**CONCLUSION**

Raman spectroscopy is a powerful technique for the study of nanotube-based systems. Precious informations can be obtained on the structural and electronic properties of the nanotubes, whether they are in the solid state or dispersed in a liquid. The frequency of the radial breathing modes can be used to estimate the diameter of the nanotubes providing that the effect of molecular pressure of the solvent is accounted for. Charge transfer between nanotubes and donors or acceptors can also be probed both in the solid and liquid states.

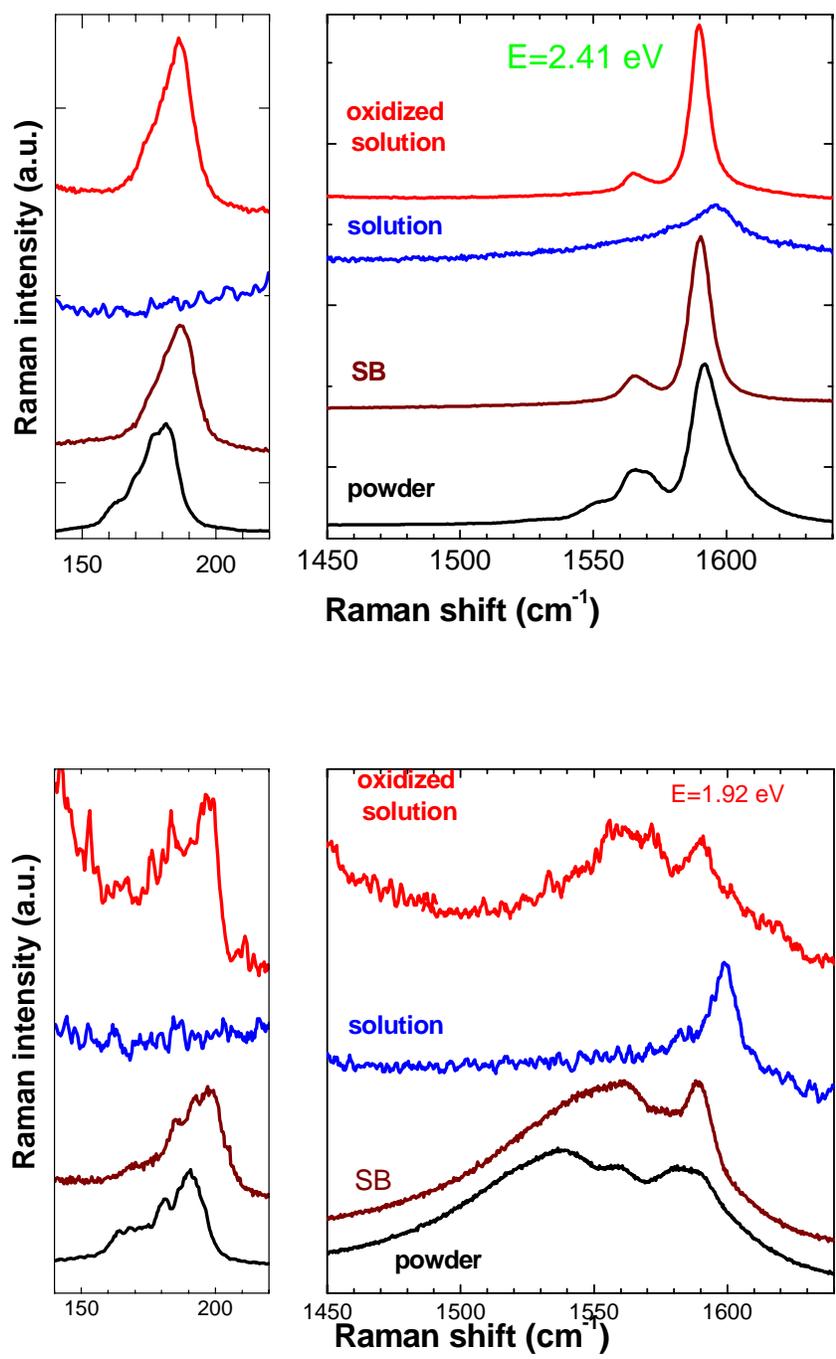

**Figure 3** : Compared Raman spectra for various nanotube samples : powders, suspensions (using Triton X100), and solutions, for laser energies 2.41 eV (top) and 1.92 eV (bottom).

**ACKNOWLEDGEMENTS**

We aknowledge fruitful discussions and collaborations with D. Riehl at DGA, Arcueil, P. Poulin at CRPP, Bordeaux and P. Petit at ICS, Strasbourg.